\documentclass[]{article}
\usepackage{setspace}
\usepackage{fullpage}
\usepackage{hyperref}
\usepackage{multicol}
\usepackage{natbib}
\usepackage{enumitem}
\setlist{nolistsep}
\newenvironment{itemize*}%
{\begin{itemize}%
  \setlength{\itemsep}{0pt}%
  \setlength{\topsep}{0pt}%
  \setlength{\parsep}{2pt}%
   \setlength{\parskip}{0pt}}%
{\end{itemize}}

\newenvironment{enumerate*}%
{\begin{enumerate}%
  \setlength{\itemsep}{0pt}%
  \setlength{\topsep}{0pt}%
  \setlength{\parsep}{2pt}%
    \setlength{\parskip}{0pt}}%
{\end{enumerate}}

\title{How to Scale a Code in the Human Dimension}
\author{Matthew J.~Turk (\texttt{matthewturk@gmail.com})\\Columbia University}
\date{}

\begin{document}

\maketitle

\noindent\makebox[\textwidth][c]{%
    \begin{minipage}{0.8\textwidth}
\noindent\hrulefill\par

\textit{This is a re-telling of a talk given at Scientific Software Days in
December, 2012, at the Texas Advanced Computing Center.  The slides and video
of this talk can be found at \url{http://scisoftdays.org/meetings/2012/}.}
\\ \\
\textbf{Abstract:} As scientists' needs for computational techniques and tools
grow, they cease to be supportable by software developed in isolation.
In many cases, these needs are being met by communities of practice, where
software is developed by domain scientists to reach pragmatic goals and satisfy
distinct and enumerable scientific goals.  We present techniques that have been
successful in growing and engaging communities of practice, specifically in the
\texttt{yt} and Enzo communities.

        \hrulefill\par
    \end{minipage}}
\vspace{0.5in}
\begin{multicols*}{2}


\section{Why ``Community?''}

Astrophysics, and particularly computational astrophysics, is dominated by
vertically-integrated, small-population research collaborations.  The concept
of a community of researchers -- sharing physics modules, improvements to
simulation codes, analysis techniques, technology -- is somewhat foreign.  In
fact, this idea of ``community'' is often viewed as a \textit{detriment} to the
individual researcher, rather than as a benefit to the field.  I participate in
two vibrant, active communities in computational astrophysics, those
surrounding the simulation code Enzo and the analysis code \texttt{yt}.  Within
these two communities, we have found a somewhat surprising result: \textbf{the
empowerment of the community does not come at the expense of individual
success.}  In fact, with actively shepherded and cultivated community
participation and processes, the opposite is true: the betterment of the
community comes at the enrichment of the individual.

In addition to the concrete, measurable benefits we receive from
community-focused development, we have also realized that \textbf{community
development is essential to the continued health of the field of computational
astrophysics.}  By focusing on developing a community of \textit{practice},
where the goals and technology are driven by active participants, we have been
able to expand the class of astrophysical problems to which the technologies
have been applied.  This has led not to consolidation of research interests,
but rather a broadening, with improvements, technology, and techniques being
directly shared between working domain scientists.

In this paper, I outline the infrastructure and the techniques with which we
have cultivated these two complementary, but distinct, communities and the
various conscious decisions we have made to ensure their growth and
sustainability.

\section{Introduction: \texttt{yt} and Enzo}

The \texttt{yt} Project (\url{http://yt-project.org/}) is an open source,
community-developed analysis code for simulated astrophysical data.  Largely
written in Python, Cython and C, it is parallelized using MPI and OpenMP and
has been used to analyze datasets whose size range from small (tens of
megabytes) to large (tens of terabytes).  \texttt{yt} is designed to abstract
out underlying technical aspects of simulation data such as file format, units,
geometric conventions, and parameter storage in such a way that is neutral to
the underlying simulation platforms.  The flagship simulation platforms, where
we conducted detailed testing and support all functionality, are are Enzo,
FLASH, Orion, Castro, Nyx, Piernik and NMSU-ART.  In the current development
branch (described briefly below) this has been expanded to include limited
support for SPH and N-body codes such as Gadget, as well as full support for
Octree codes such as RAMSES and ART.  \texttt{yt} provides a language for
describing physical regions and applying data processing techniques to those
regions: rather than focusing on selecting grid patches, particles or octs and
then masking out overlapping regions, it develops concepts of geometric
regions, regions defined by fluid quantities, and processes that transform
those regions into quantitative values.  \texttt{yt} is best thought of not as
an application, but as a library for asking and answering questions about data.
This aspect of \texttt{yt} naturally encourages technical contributions, as
every user of the software package typically writes analysis code that builds
upon its underlying machinery.  Although its mission has expanded in recent
years to include the application of microphysical solvers, standardized
input/output data formats, creating initial conditions, and even studying data
from earthquake simulations, at its core \texttt{yt} remains a tool for
interrogating astrophysical data and answering questions about the underlying
physics of that data.

The defining characteristic of \texttt{yt} is not that it is written in Python,
or the operations it does, but rather that it is supported by a participatory
community of scientists interested in both using and developing it.  The
\texttt{yt} community draws its members primarily from computational
astrophysicists: individuals who conduct and analyze simulations.  The first
version was written in late 2006 by and for a single scientist, but over the
last several years has attracted 170 subscribers to the general mailing list
and 50 for the development mailing list; in 2012 it was identified as one of
the most highly used codes on the NSF NICS analysis machine Nautilus
\citep{6203491}.  Over the course of its history, nearly 40 people have
contributed changesets, ranging from tiny to very large, and the mailing lists
are relatively active, averaging between one and four messages a day.  In 2012
we held our first workshop at the FLASH center in Chicago, and we are holding a
second workshop in March of 2013 at UC Santa Cruz.  Despite this relatively
high level of activity for a project in computational astrophysics, the
community has faced several challenges as well as taken steps to directly
address these challenges.

I am the original author of \texttt{yt}, having created it during my graduate
school career to analyze and visualize data created by the simulation code Enzo
(\url{http://enzo-project.org/}), an adaptive mesh refinement simulation code.
Enzo itself has undergone a number of changes over its years, both in the
manner in which it is developed and the problems to which it can be applied.
Enzo began as a codebase largely developed by a single individual, but through
the stewardship of the Laboratory for Computational Astrophysics (LCA) it grew
into a large, open source (but mostly not community-based) development project.
A direct consequence of both the technology and the prevailing mindsets led to
divergent lines of ``public'' and ``internal'' development, resulting in a
highly fragmented community of users and developers and many different branches
of the code itself.  Over the course of a few years in 2009 and 2010, through
concerted efforts from the Enzo community, it transitioned from a closed
development model with periodic open releases into a fully-open, transparent
development model based around contributions from community members.  This
included resolving many divergent code bases and produced the Enzo 2.0 release.
The code is now developed using many best practices of software development
including testing, version control, peer review, infrastructure design and
investment of stakeholders in the development roadmap.  This has resulted in
contributions of both physics modules and infrastructure improvements, and has
even brought to light bugs that have subsequently been fixed by community
members.  This transition, from semi-open to community-driven, has brought an
energy and excitement to the development of Enzo that seems likely to sustain
it for years to come.  Many of the items discussed below, about how to shape
and foster community, have been applied to the Enzo community as well as
\texttt{yt}.

\section{What is ``Community''?}

Open Source is often used as a synonym for community development, but in
practice they are better thought of as two different, but overlapping, modes of
development.  The growth of a community, where individuals participate in
discussions, report problems, provide enhancements, and support other community
members, is neither necessitated by software being open source, nor is it a
foregone conclusion for open source software.

For scientific software, this situation is slightly more complex, as the
adoption of open source methods are often met with resistance.  In addition to
this, members of a given community related to software are just as likely -- if
not more likely -- to be working on similar projects, competing for mindshare
among the academic public, and even competing for funding or jobs.  Building
communities that are able to thrive despite these barriers can result in
considerable, non-local benefits: the aphorism ``a rising tide lifts all
ships'' is nowhere more true than in community-driven scientific codes.  Even
the presence of other, engaged community members means that there are more
people able to answer questions from newcomers and provide assistance and
energy toward solving problems.

The concept of openness in science amongst academics is something of a paradox.
Often, utilizing commodity tools is viewed as a very positive trait, whereas
releasing software is occasionally viewed with skepticism.  While this is
changing, and in some ways quite rapidly, the idea of source code being shared
between potential competitors is still often seen as dangerous or even anathema
to scientific progress.  This typically breaks along three primary
objections:\footnote{For a broader and more quantitative study, see
\cite{citeulike:8795002}}

\begin{enumerate*}
\item Why should I give up my competitive advantage?
\item How can I manage supporting a code?
\item What if someone finds a bug in my work?
\end{enumerate*}

Answering these questions in detail is beyond the scope of this document.  But,
a few comments can be made.  \textbf{The first two objections to releasing code
are directly addressed by a community of practice.}  As noted above, the
benefits from collaborative, participatory communities can alleviate the
burdens of support as well as provide advantages to scientific inquiry that
would otherwise not be present.  In both the Enzo and \texttt{yt} communities,
as the community has scaled in size, with it has scaled the number of eager,
helpful participants in contributing modules (which are then shared,
collectively) as well as provide support for problems and issues on the mailing
list.  When discussing these aspects of competitive advantage and support, it
is also worthwhile to frame the discussion in terms of generalization and
specialization.  As an example, in the early stages of developing a simulation
code, it's entirely reasonable that a single individual can manage every aspect
of the code: the IO framework, parallelism, hydrodynamics, gravity, and so on.
But as the simulation code becomes larger, more general, and more mature, it
becomes unwieldy for a single person to manage and develop all aspect of a
code.  Within Enzo we have seen this as the code has grown to include many
different hydrodynamic modules (including magnetohydrodynamics), chemistry
modules, radiative transport, star particle implementations, and parallelism
strategies.  The overall generalization of the code base has resulted in a deep
specialization in some areas of the code.  By investing in a community of
individuals, the overall management costs are reduced as individuals specialize
in different subsections of the code.  This pipeline for conducting research
using a code such as Enzo shares many characteristics with observational
research; the data is constructed using the simulation platform, relies on many
pieces of diverse infrastructure such as IO libraries, libraries for
parallelism and libraries for solving differential equations, and is then
passed on to the analysis platform which processes the data to produce results.  

The third objection enumerated above is somewhat more subtle.  The implicit question
is not, ``What if someone finds a bug?'' but really ``What if someone sees my work
and is able to show that it is flawed?''  This objection is the most challenging, as
it neatly aligns with the conflict between personal advancement and the collective
advancement of science.  The glib, simple solution to this would be to prioritize an
incremental, layered approach to science rather than preserving the professional
benefits of previous, potentially incorrect results.  In this approach, ``bugs'' and
flaws in results are not hidden from view and defended from exposure, but rather
found and corrected.  Unfortunately, such prioritization is a subtle form of the
prisoner's dilemma: it is maximally successful when all members of the community
participate in revealing shortcomings and problems.  The personal solution is
somewhat more subtle, and not entirely clear.  From a pragmatic perspective,
identification of bugs enables higher-quality, longer-lasting results, less likely to
be overturned by subsequent investigations by competitors.

As computational science matures, and as computational techniques permeate
every aspect of scientific inquiry, it is natural that the software utilized in
scientific inquiry grows more complex.  Scientific software in many respects,
particularly in astrophysics, is thought of as something of a second-class
citizen -- in years past, the concept of scientific software being developed in
isolation, placed on a website and largely abandoned became pervasive.  This
methodology simply will not scale with the complexity of projects necessary for
modern scientific inquiry; we have reached the age of advanced algorithms being
applied in non-trivial ways to complex, physically-rich datasets.  The
cyberinfrastructure necessary to address problems in computational science is
no longer tractably solved by individuals working in isolation -- community
projects must become the new norm.

Part of the reason that communities developed around scientific codes can be
beneficial is also a component of the challenges within these communities.  An
active community with participants sharing enhancements, features, and
assistance relies on the participants developing those enhancements and
understanding the code base well enough to provide assistance to less
experienced individuals.  Scientific code development is strongly driven by
pragmatic, short-term needs of scientific inquiry; as a result, most
improvements are motivated by the next paper, or the next talk, or advancing a
particular line of inquiry likely to result in a publication and the affiliated
respect from peers for solving a challenging scientific problem.  The common
argument against sharing advancements, techniques and tools is that it
undercuts competitive advantage.  This form of individualism results not
only in researchers refraining from sharing, but more importantly, prevents
them from benefiting from the work of others.  A strong community of sharing
results in a stronger technical code base, even if it has the side effect of a
moderate reduction in perceived intellectual priority for individual
researchers.  If I share a technique that can be applied to Population III star
formation simulations, I can no longer be the only one to apply that technique
-- and thus the only person who can answer that type of problem.  I lose a
small amount of intellectual priority.  However, I am now able to receive
improvements to the technique, suggestions for further enhancements, and am thus able
to extend my scientific inquiry further than before.  The most important aspect of
this is that altruistic goals (sharing, reproducibility, openness) align exactly with
non-altruistic goals.  It does not matter if I share because I hope to gain something
in return or because I believe in openness; the end result is that the \textit{health
of the community} has been improved by participation.

The strongest bias we have seen in \texttt{yt} is not against releasing of code
or contributing back, but against the notion of conducting any development at
all -- whether that be contributing code, documentation or even reporting bugs.
This cognitive block comes from a fundamental misunderstanding of how
scientific codes are developed, one that we have been attempting to remedy
within our own ranks.  The traditional view of open source scientific code
development has been that of two groups -- developers and users.  This
segregation of individuals results in an implicit, but difficult to overcome,
feeling of boundaries between responsibilities.  Often responsibilities such as
verification of results, inspecting results, and tracking (and understanding)
modifications to the code base are left to ``developers.''  Unfortunately, all
of these responsibilities are essential components of the scientific method!
Segregating responsibilities in this way leads to misunderstandings about the
nature of scientific codes, many of which are constantly under development and
improvement, and the nature of the results that they can produce.  This is even
evidenced in simple things such as using the word ``user'' to describe
community members, further emphasizing a distinction between people who ``use''
the code and people who ``make'' the code.  In \cite{6223004}, the authors
thoughtfully describe the role of scientists in software not as a black and
white, user or developer distinction, but rather as a continuum of greys.
Often, the self-application of a term like ``user,'' with its connotations,
results from self-assigned roles, or perceptions of individual abilities.  As
peers in a global scientific community, \textbf{the distinction between
``users'' and ``developers'' is actively harmful}, when in reality scientists
are tasked with occupying that grey area in the middle.  

In both the \texttt{yt} and Enzo communities, we have seen this behavior play
out numerous times.  Those members of the community who have been the most
giving of enhancements, assistance, and technology (scripts, modules, bug
fixes) are typically those who are able to best take advantage of the
technology to ask complex questions about the physical world.  Furthermore, the
process of opening up a module and contributing it to an open source software
community provides the opportunity for receiving enhancements in functionality,
bug reports, and synergistic applications of that module.

In the \texttt{yt} project, there are a number of individuals who are the
primary contributors and a larger number of individuals who have provided
occasional bug fixes or individual, isolated modules.  The influence of
non-developing community members on code changes, roadmap efforts, and
development methodology is not only allowed, but encouraged and
\textit{solicited}.   Furthermore, we actively solicit contributions in the
form of scripts, modules, and enhancements or modifications to the \texttt{yt}
codebase.  This initiative has resulted in several positive effects, all of
which have strengthened \texttt{yt} as a \textit{community}, independent of
their effects on \texttt{yt} as  \textit{software}.  Individuals who do not
feel motivated to or capable of contributing in a technical sense are more
likely to contribute through helping other users, providing documentation, and
even networking and advertising the project.  But the more prominent effect is
a sense of investment in the project.  This sense of investment results in
quantifiable results (such as more mailing list activity, a larger number of
commits) as well as results that are more difficult to quantify, such as
positive feelings, friendly discourse, excitement, word of mouth, and so on.

\section{Challenges}

Even beyond these concerns, community building can be challenging within the
confines of the traditional realm of academia.  The reward structure in
astrophysical academia, for funding, jobs and mindshare, is highly-correlated
with influence, high-impact papers, and citation counts\citep[see
also][]{DBLP:conf/cscw/HowisonH11}.  Often developing software, or
participating in academic communities, is seen as orthogonal to these goals,
even though this is not necessarily true.  In light of this, soliciting
development contributions is particularly difficult.  Developing a tool in
support of a publication (for instance, a module for \texttt{yt}) is seen as a
worthwhile use of time.  Despite the benefits that come from making that tool
available, the time needed to ``clean it up'' and provide a modicum of support
is not as immediately beneficial, as it does not result in an additional
publication, additional citations, and so on.  Often, those contributions that
would be the most beneficial to the project as a whole are the most difficult
to motivate.  These secondary, yet important, goals include improvements to
documentation (particularly narrative, instructive documentation), testing of
individual components (as opposed to large-scale integration testing), and
infrastructure improvements such as optimization and maintainability
refactoring.  As a result, tool builders are not always favored by the academic
reward structure.  This is not unique to computational astrophysics, but is
also seen frequently within the instrumentation community.  A common pattern
seen within \texttt{yt} is a spirited discussion of idealistic goals for the
project, but then pressing deadlines and other local concerns typically temper
enthusiasm and execution.  I myself am not only guilty of this, but perhaps the
most common instigator!

The challenge to community building I find the most worrisome is that presented
by the so-called ``citation economy'' in astrophysics.  The influence of a
given piece of code is often measured by citations to a method paper, which is
by definition a fixed and archived document.  In 2011, a method paper for
\texttt{yt} was published.  Enzo's method paper is still under preparation, but
papers in 1997 and 2004 are often cited as references for Enzo.  Because ADS
provides reverse-indexing capabilities, the number of citations to these papers
provides an immediate method of gauging the influence of the software that they
describe.  In the intervening time between publication of the method papers,
new contributors have joined the collaborations and contributed substantial and
non-trivial enhancements to the code base.  In fact, this is not only what has
already happened, but an explicit, named goal of the future of these two code
bases!  This results in an unfortunate conflict of interests between new
contributors and existing contributors.  It is in the best interests of
existing contributors, who contributed to the code base prior to the
publication of a method paper, to consolidate citations in the canonical paper;
unfortunately, it is in the best interests of new contributors to publish a new
method paper, so that they can begin to ``collect'' citations.  In whose
interest is it to contribute to a code base without rewards mediated through
citations, one of the most common metrics of success in academia?  I believe
this is among the greatest challenges to community developed codes in
astrophysics, one that will continue to grow in importance as software projects
inevitably scale beyond a handful of contributors\footnote{Alternate metrics of
software citation are being explored by, e.g. the ASCL
\citep{2012ASPC..461..627A}, but they primarily address discoverability of
codes and do not attempt to address the issue of distributing credit to
contributors.}.


\section{Strategies}

Not all of the challenges enumerated above can be addressed directly.  However,
in both the \texttt{yt} and Enzo communities, we have addressed the technical
and social challenges through conscious development of infrastructure and
techniques.  These have been designed to foster good will, encourage
contributions, and build social capital between contributors.  Below, I list
several steps we have taken over the last several years toward the goal of
growing and engaging communities of practice.\footnote{A related work is
\citet{citeulike:11831265} where the authors have compiled a set of very useful
and helpful guidelines for scientific software development.}

Most important, however, these need to be viewed as the investment they are.
These strategies, particularly the social strategies, require not only focused
energy and time, but an emotional investment.  \textbf{You must design the
community you want.}  This design extends far beyond designing software and
algorithms; it includes thinking about the diversity of contributors and
community members, the tone of discourse within the community, the projected
enthusiasm within the project, and even the congeniality with which feedback --
especially critical feedback -- is received.  The culture seeded in a community
will self-propagate; whether this is a culture of neglect, a culture of
homogeneity, a culture of kindness, a culture of brash arrogance or even a
culture of openness, this will flow outward from the
core\citep{citeulike:7888211, Trapani:2011, Allsopp:2012}.  Both Enzo and
\texttt{yt} have attempted to build cultures that promote respect and
excitement.

\subsection{Technical Infrastructure}

Within both \texttt{yt} and Enzo, we utilize several pieces of technical
infrastructure that ease the process of growing communities \citep[see
also][]{citeulike:478633}.  These include open and freely-joinable mailing
lists, a completely open development process based on the distributed version
control system (DVCS) mercurial, and a code review and mentoring process to
ensure contributions are high-quality, with a minimum of friction.
Furthermore, we utilize methods of communication that enable participation and
that are tuned for low- to high-latency interaction.

Both \texttt{yt} and Enzo have seen an enormous growth in contributions
following migration to DVCS; this is not unique to these projects, but is a
hallmark of the lowered barrier to entry presented by DVCS.  In contrast to
centralized version control systems such as CVS and Subversion, DVCS systems
are fully-distributed.  Every clone, or checkout, of a repository is a peer
with every other clone; this enables anyone who checks out a repository to
track their own changes to the repository, and provides the (technical) ability
to much more easily contribute changes upstream.  The unique versioning of
local copies of a code base also allow unique revision specifiers to be applied
globally; rather than ``version 4.2 with modifications (unspecified)'' code
that generates a given result can be uniquely identified with a globally-unique
hash.  Both \texttt{yt} and Enzo are hosted on BitBucket, a code hosting
provider (roughly functionally equivalent with others such as RhodeCode,
GitHub, Gitorious and Savannah).  It provides technical infrastructure for code
review, accepting changes, tracking contributions and identifying specific
revisions of the code.  By using this system we lower the barrier to entry for
contributors, enabling submission of locally-developed changes.  By using DVCS,
we encourage scientists using \texttt{yt} or Enzo to track their \textit{own}
changes to it.  Even if this code is never submitted for inclusion in the
primary repository, this enables provenance tracking and debugging.

Development of scientific software also requires several difference degrees of
communication; as the software itself is not the only project contributors are
working on, we have found that communication is naturally divided into three
categories, based on latency and urgency.  Immediate communication, such as
in-person meetings (or ``code sprints'') or Google Hangouts, are useful for
low-latency planning, discussion, and collaboration.  Tools such as Internet
Relay Chat (IRC) provide a means of medium-latency communication; often in the
\texttt{\#yt} channel on FreeNode, an IRC network, between 10 and 20 people can
be found at any time.  Messages left here are usually responded to relatively
quickly, which leads to a low-latency discussion where problems can be resolved
much more quickly than through email.  Finally, nearly all planning and
detailed discussion happens through high-latency mediums such as comments on
code changes and mailing lists.  The mailing lists for \texttt{yt} (which are
titled the unfortunately-chosen names \texttt{yt-users} and \texttt{yt-dev})
are open, indexed by Google, and freely joinable.  We have often (successfully)
encouraged discussions here to turn into collaborations and code sharing, and
in fact we have had a number of successful opportunities for long-term planning
and invasive code changes here.  They serve not only as broadcast media, but as
venues for soliciting input and contributions.

Nearly as important as providing mechanisms for communication is reducing the
barrier to entry for new contributors.  This includes ensuring a smooth process
to finding the appropriate location in the source code, making changes,
submitting those changes for review, and a kind, thoughtful mentoring process
for new code contributors.  The most common path we have seen for new code
contributors is relatively straightforward:

\begin{enumerate*}
\item Individual applies software package to meet own goals
\item Individual develops a modification to the code, for either an
enhancement, a bug fix, or a contribution of an example
\item This change is submitted for inclusion in the main \texttt{yt} repository
\item The change is read, reviewed and tested by community members to ensure
that standards for coding, documentation and testing are all met.
\item The individual begins to participate in the community
\end{enumerate*}

The main codebase is not suitable for contributions of all types; we have found
individuals eager to contribute scripts that exist as supplemental tools, or
that were used in the writing of a paper.  To support this desire, we provide a
location (\url{https://hub.yt-project.org/}) to submit these scripts as well,
where links to external source code repositories are collected alongside
descriptions of the individual projects.  In this way, by providing mechanisms
and structures, as well as an open and active solicitation for technical
contributions, we have found that we can foster a sense of community and
accomplishment among individuals.

We have found that even minor barriers to contribution (or even software
deployment!) can build up a ``technical friction'' to participation that
results in losing contributions and community members.  In an effort to
alleviate this, we provide easy access to the source code, detailed
instructions for contributing code, a mechanism for communication, and scripts
that assist with the entire process.  We have found that by ensuring that we
have a uniform review process (where even code written by founding members of
the community is reviewed) we ensure that the code is of acceptable quality,
tests are included, and that undocumented code is discouraged.

\subsection{Social Techniques}

In \citet{opac-b1134063} the authors describe a strategy for social
interactions among so-called ``geek'' teams.  I will not provide a detailed
retelling of this, but despite focus on applying the techniques described in
the book to software engineers, in many ways they can be applied directly to
collaborations around scientific software projects.  The authors enumerate
three characteristics which they suggest applying to all communication and
interaction:

\begin{itemize*}
\item Humility
\item Respect
\item Trust
\end{itemize*}

Communication by text, in particular about technical topics, is often stripped
of the nuances and inflections that convey emotion.  In textual media,
therefore, it is even more important to guide discussion in a way that
encourages participation.  Within the \texttt{yt} community, we are very
careful to ensure that the tones on the mailing list, code review and in IRC
should be conducted in this way: the health of a community can be judged by how
it treats newcomers and novices as well as how it treats experienced, advanced
participants.  Even in framing analysis interactions between peers in this way,
one can see how communication can serve the growth of scientific communities.
In scientific software communities, an aspect that often goes unmentioned is
that we arguably want to foster discussions between \textit{peers}, not
discussions between an elite class and a subservient class.  By guiding the
discussion with a focus on humility, respect and trust (HRT as abbreviated by
Fitzpatrick and Collins-Sussman), discussions can become more congenial, more
productive, and can lead to a greater spirit of collective focus.

\textbf{Humility} can be well-characterized in how a community responds to
negative feedback.  In a community in which I participate, I witnessed a
discussion between an experienced community member and a newcomer.  The
newcomer reported what they thought was a bug in a particular routine.  The
experienced community member, who had made the change that introduced the
potential bug, responded with a very abrupt, dismissive response.  ``It's like
that for a very good reason.  Don't touch it.''  By terminating discussion at
that point, it sent the clear signal that a discussion of the reasoning behind
a code change was simply not necessary; the message was not only that the
reasoning was beyond reproach, but that a discussion was not worth the time it
took to have.  The better solution, of an open and humble discussion of the
background and reasoning behind the decision, would have been to engage the
newcomer and explain the reasoning behind a decision.  This takes an investment
of time, but certainly no more than the investment of time it would take to
justify such a decision to a referee or journal reviewer, and the potential
gains in this case would be a new contributor and an increase in social capital
amongst the community.

\textbf{Respect} can be seen in how peers contribute to the development of the
code.  In principle, when developing a scientific code, we are engaging in
scientific discourse, attempting to push the field of astrophysics forward in
both our understanding of and our ability to ask questions of the natural
world.  One of the most common ways I've seen this ignored is when a person
writes to a mailing list or appears on IRC and says some variation of ``I've
noticed something is acting strangely with ...''  The all-too-common response
is simply, ``You're doing it wrong,'' or even worse, a dismissive instruction
to ``Just read the documentation.'' This is rarely even prefaced with the
obvious question of, ``How did you expect it to act?''  It terminates
discussion and discourages individuals from returning.  This has the direct
effect of reducing participation, and disproportionately impacts new
contributors.  A lack of respect can infect and transform a community into an
unfriendly, unwelcoming environment; the most valuable attribute a community of
scientific software developers has is its ability to engage in scientific
practice, and that is obstructed by a lack of respect.

\textbf{Trust} within scientific software communities is perhaps more subtle.
What we have seen with \texttt{yt} and Enzo is \textit{not} that trust is an
unwavering faith in someone to produce high-quality, scientifically correct
code, but a trust in the process and the stewardship of a code.  Within
communities of scientific software, we have seen this evidenced when
contributors \textit{let go} of a project.  As a community matures, individuals
cannot contribute to the breadth of sub-projects that they did when it was
young; furthermore, as scientific interests shift, the leader of an individual
project or aspect of the scientific code may move on to other things.
\textit{Trusting} that the community will be able to steward and shepherd that
project is a crucial ingredient in addressing sustainability and burnout, and
an essential ingredient in expressing trust for peers and other community
members.

The motion of projects between individuals is a difficult social situation.  We
attempt to address this in \texttt{yt} by emphasizing \textit{pride} over
\textit{ownership}.  \textbf{By changing the conversation from dominion to
stewardship, this changes how individuals approach projects, and helps enable
them to regard external contributions in a positive way, rather than as threats
to their accomplishments.}

HRT are not always easy attributes to focus on.  Particularly in a competitive
field such as astrophysics, it can be difficult to spend time thinking about
how words or conversations will be perceived, particularly when these
conversations are between potential competitors.  But by remaining focused on
engaging community members as peers, treating them with HRT and spending time
and energy on thoughtful communication, the community will be more likely to
grow and flourish.

\section{Conclusions}

As computational science has grown both more complex and more pervasive,
communities of practice surrounding scientific software projects have become
essential to the health and vibrancy of computational science projects.
Unfortunately, the academic reward system does not always align with the
development of software projects, which can lead to stagnation or corner
cutting in important tasks such as infrastructure, documentation and testing.
This largely results from the overlap between the traditional roles of
developers and users in software development, a distinction that can in fact
actively harm the scientific process.

To grow communities around any software project, the structure and character of
the community itself must be carefully considered; you must \textit{design}
this community.  This includes setting the tone of discourse, consciously
fostering a diverse set of participants, and being open and enthusiastic.
This process can be assisted with technical infrastructure.  This includes
using distributed version control systems, conducting open communication in
media that suit the style of communication, and streamlining the participation
process and providing a clear method of contributing.  Technical infrastructure
alone cannot ``solve'' the development of a community; social infrastructure
and standards of conduct must also be developed.  Conducting business while
remaining focused on humility, respect and trust can help to ensure that
newcomers feel welcomed, existing contributors and peers feel validated and
respected, and that ultimately as interests change over time projects do not
stagnate under the weight of code or project ownership.

A vibrant, active community brings sustainable development, synergistic
applications and development of code, and considerable rewards in potential
collaborations and discourse.  I am grateful for the opportunity to have
participated in both the \texttt{yt} and Enzo communities, and grateful for the
concrete rewards -- scientific, social and technical -- that my participation
has allowed me.

As a closing note, despite all of the successes we have seen in the \texttt{yt}
and Enzo communities, we still have considerable room to grow from both the
social and technical standpoints.  I hope that as communities driven both by
long-reaching goals and the pragmatic needs of scientific inquiry, we can
continue to remain healthy, vibrant and growing.  I believe that the single
biggest problem within scientific software communities is ensuring credit for
community participation can be shared with new contributors.  This affects
motivations at essentially every level, and because of its very nature can
dramatically affect the health and stunt the growth of scientific software
communities.

\section*{Acknowledgments}

I would like to thank the organizers of the Texas Scientific Software Days
conference, Victor Eijkhout, Sergey Fomel, Andy R.~Terrel, and Michael Tobis for a
wonderful conference full of excellent talks.  I'd also like to thank the other
attendees and speakers for fascinating and enlightening conversations about
scientific software.  I am supported by an NSF CI TraCS fellowship (OCI-1048505).
Any opinions, findings, and conclusions or recommendations expressed in this
publication are my own and do not necessarily reflect the views of the National
Science Foundation.

I would also like to thank the members of both the \texttt{yt} and Enzo
communities.  Much of this talk -- and approach to community building -- was
developed in collaboration with them.  I'm grateful for feedback provided by
Nathan Goldbaum and Brian O'Shea on a draft of this manuscript.  In addition, I
would like to thank Britton Smith and Greg Bryan both for extremely valuable
feedback in preparing this paper and for leading by example in community
development for scientific codes.

\bibliography{refs}

\end{multicols*}
\end{document}